\documentclass[preprint,12pt,number]{elsarticle}
\usepackage{amsmath,amssymb,amsfonts}
\usepackage{algorithmic}
\usepackage{graphicx}
\usepackage{textcomp}
\usepackage{float}
\usepackage{mathtools}
\usepackage{array}
\usepackage{url}
\usepackage{multirow}
\usepackage[fixed]{fontawesome5}
\usepackage{xurl}
\usepackage{longtable}
\usepackage{hyperref}
\hypersetup{
    colorlinks=true,
    anchorcolor=black,
    citecolor=blue,
    linkcolor=blue,
    filecolor=blue,      
    urlcolor=blue,
}
\usepackage{amssymb}
\usepackage{amsmath}
\usepackage{lineno}

\journal{To be determined}

\begin{document}

\begin{frontmatter}



\title{A survey on the impact of emotions on the productivity among software developers} 

\author[inst1]{Pawel Weichbroth}
\author[inst2]{Maciej Lotysz}
\author[inst1]{Michal Wrobel}

\affiliation[inst1]{organization={Department of Software Engineering, Faculty of Electronics, Telecommunications and Informatics, Gdansk University of Technology},
            addressline={Narutowicza 11/12}, 
            city={Gdansk},
            postcode={80-233}, 
            state={pomorskie},
            country={Poland}}

\affiliation[inst2]{organization={Gdansk University of Technology},
            addressline={Narutowicza 11/12}, 
            city={Gdansk},
            postcode={80-233}, 
            state={pomorskie},
            country={Poland}}

\begin{abstract}
\textbf{Context:} 
The time pressure associated with software development, among other factors, often leads to a diminished emotional state among developers. However, whether emotions affect perceived productivity remains an open question.

\noindent \textbf{Objective:} 
This study aims to determine the strength and direction of the relationship between emotional state and perceived productivity among software developers. 

\noindent \textbf{Methods:}
We employed a two-stage approach. First, a survey was conducted with a pool of nine experts to validate the measurement model. Second, a survey was administered to a pool of 88 software developers to empirically test the formulated hypothesis by using Partial Least Squares (PLS), as the data analysis method.

\noindent \textbf{Results:}
The results of the path analysis clearly confirm the formulated hypothesis, showing that the emotional state of a software developer has a strong positive, and significant impact ($\beta = 0.893$, $p < 0.001$) on perceived productivity among software developers.

\noindent \textbf{Conclusion:}
The findings highlight the importance of managing and improving developers’ emotional well-being to enhance productivity in software development environments. Additionally, interventions aimed at reducing burnout, stress, and other negative factors could have a considerable impact on their performance outcomes.

\end{abstract}


\end{frontmatter}


\section{Introduction}
\label{sec:introduction}
In general terms, productivity is a measure of output relative to input \cite{keh1997classification}.
More specifically, productivity is defined as the efficiency with which resources are used to produce goods and services. Among all the available resources, humans are considered one of the most valuable assets \cite{fulmer2014our}, playing a particularly essential role in mentally demanding workplaces \cite{rajgopal2010mental}. From an individual perspective, engaging in mental work can provide a sense of accomplishment and purpose \cite{dupuis1995examination}. However, it can also result in heavy workloads, time constraints, and cognitive demands. These factors can lead to stress \cite{bowling2012workload}, burnout \cite{zanabazar2022relationships}, and decreased well-being \cite{gaeta2020need}.

From a business perspective, mental health in the workplace has significant and far-reaching consequences. An estimated 12 billion workdays are lost to depression and anxiety worldwide each year, resulting in an annual loss of productivity of US\$1 trillion \cite{WHO2024}. Some estimates suggest that major depression alone costs employers between \$31 billion and \$51 billion per year in lost productivity \cite{Taylor2020}. The projected cost of mental health conditions and their related consequences is expected to exceed \$6 trillion by 2030 \cite{WEF2025}. Unsurprisingly, interest in managing emotions in the workplace is on the rise among organizations \cite{morris1997managing, resendiz2021novel, nawaz2024nexus, yadav2025emotion}.

In a natural way, emotions are an inherent part of the human experience during work performance \cite{barrett2006solving, kowal2025fire}. Like any other white-collar worker \cite{kowal2024organizational}, software developers perform office-based intellectual work including designing, developing, testing, and maintaining software applications. As practice shows, they often have to work under deadline pressure \cite{van2018under}, demonstrate the ability to prioritize tasks \cite{diamantopoulos2023semantically}, and take full responsibility \cite{kolimar2022delivery}. In addition, as software development projects become more complex \cite{kaufmann2022does}, developers must not only maintain high-quality code and collaborate effectively \cite{bani2014collaborative}, but also preserve productivity to meet time constraints and deliver the expected functionality \cite{mishra2023structured}. All these factors can trigger strong emotional responses.

Prior studies have highlighted that software developers are more prone to anxiety \cite{graziotin2017consequences}, burnout \cite{singh2012antecedents}, fatigue \cite{palacio2022profession}, and stress \cite{meier2018stress} than those in other office jobs. Although emotions play a vital role in the workplace, little recent research has examined their relationship with software developer productivity. To date, research directly addressing this issue remains limited. Therefore, this study seeks to fill this gap, motivated by the need to further explore this phenomenon.

With this in mind, our study offers three important contributions.
First, it re-operationalizes the measurement constructs of emotional state and perceived productivity. The underlying foundations are rooted in the latest research and are validated through qualitative, survey-based research conducted with a pool of experts.
Second, it confirms the existence of a strong, positive relationship between emotional state and perceived productivity. Third, it emphasizes the importance of implementing appropriate measures to mitigate emotional risks and maintain a healthy work environment.

The remainder of this paper is organized as follows. 
Section~\ref{sec:related-work} presents related work carried out at the intersection of emotions and productivity.
Section~\ref{sec:Research-framework-hypothesis-formulation} describes the research framework and puts forward the hypothesis.
Section~\ref{sec:methodology} outlines the methodology employed.
Section~\ref{sec:results} presents the findings from the analysis.
Section~\ref{sec:discussion} examines both the theoretical and practical implications, along with the study’s limitations.
Section~\ref{sec:conclusion} concludes the paper and provides possible directions for future research.

\section{Related Work}
\label{sec:related-work}
Over the decades, the focus of software engineering research has shifted from strictly technical and process-based concerns to a broader, more human-centered approach. The evolution toward recognizing the importance of the developer's role can be traced back to pioneering works such as Gannon's Human Factors in Software Engineering~\cite{Gannon1979Dec}, which was one of the first to explicitly emphasize the significance of the individual. This perspective was further strengthened by the research of Boehm and Papaccio~\cite{Boehm1988Oct}, who argued that improving productivity and controlling software costs most effectively requires a focus on the people involved. While these efforts established the importance of developers, the study of their emotional states drew heavily from parallel advancements in organizational psychology and computer science. 

A crucial theoretical model was provided by Weiss and Cropanzano's Affective Events Theory~\cite{weiss1996affective}, which posits that workplace events trigger emotional responses that directly influence performance and attitudes. Almost concurrently, Picard~\cite{picard1997affective} founded the field of affective computing, establishing the technological basis for recognizing, interpreting, and processing human emotions. Together, these foundational contributions from human factors, organizational psychology, and affective computing created the necessary framework to investigate the emotional landscape of software development and its impact on productivity.

One of the earliest studies focusing on the emotional experiences of software developers was conducted by Shaw~\cite{Shaw2004Apr} in 2004. He applied Affective Events Theory to a study of senior-level information systems students engaged in a semester-long capstone project. Although it did not involve industry professionals, the study provided early validation of AET in the context of software development by identifying project events that triggered emotional responses. 

Nearly a decade later, this area of inquiry gained significant momentum with more detailed quantitative analyses. Wrobel cataloged the frequency and impact of various emotions and introduced the concept of ``emotional risk'' to productivity~\cite{Wrobel2013}. His analysis revealed that frustration was the most frequent and disruptive negative emotion, and enthusiasm was the most beneficial and commonly experienced positive emotion. Notably, the study also found that anger was a unique negative emotion; a significant proportion of respondents reported that anger increased their productivity. 

In a parallel study, Graziotin, Wang, and Abrahamsson employed a different methodology~\cite{Graziotin2013}. They conducted a repeated-measures study in which they observed developers working on individual projects and assessed their affective states and self-rated productivity at ten-minute intervals. Their intensive data collection provided a direct answer to their research question: happy developers are more productive. Together, these empirical studies were pivotal in moving the field from theoretical postulation to concrete evidence. They confirmed that emotions are a significant and measurable factor influencing the effectiveness of the development process, not merely a byproduct of it.

A significant body of work that advanced the understanding of affect in software engineering was contributed by Graziotin and his colleagues. Their research program systematically explored the connections between emotion and developer performance from multiple angles. Early studies provided quantitative evidence supporting the claim that happy developers demonstrate superior analytical problem-solving abilities~\cite{Graziotin2014Mar}. They also pioneered the use of fine-grained, real-time measurements, conducting repeated-measures studies that correlated the affective dimensions of valence (pleasure) and dominance (control) directly with developers' self-assessed productivity~\cite{Graziotin2015Jul}. Beyond identifying purely quantitative links, they developed a novel explanatory theory that describes how affects impact performance through concepts such as events, focus, and goals~\cite{Graziotin2015Aug}. Recognizing the methodological challenges in this new domain, they advocated for a more rigorous approach, coining the term "psychoempirical software engineering" and publishing guidelines to steer future studies away from common misconceptions~\cite{Graziotin2015Sep}. 

A major focus of Graziotin's later work shifted to the practical implications of affective states, particularly unhappiness. Through large-scale surveys, they identified hundreds of causes of developer unhappiness~\cite{Graziotin2017Jun} and cataloged dozens of distinct consequences stemming from both happy and unhappy states, creating classification schemes to guide managers and future research~\cite{Graziotin2017May, Graziotin2018Jun}. Recently, research has begun investigating the broader competency of emotional intelligence (EI) among software developers. For instance, a case study of software engineering students revealed that, although they were better at managing their own emotions than perceiving others', they developed strategies such as structured planning and peer support to enhance productivity and resolve conflicts. This highlights the significance of emotional competencies in collaborative development education~\cite{Araujo2025}.

Beyond directly surveying developers, another stream of research has sought to understand their emotional states by analyzing the digital artifacts and platforms central to their work. Murgia et al. explored this approach when they investigated whether artifacts from issue tracking systems in open-source projects carry emotional information. Their analysis of the Apache Software Foundation's issue tracker revealed that developers do express emotions such as gratitude and joy. However, the analysis also revealed a significant challenge: the more context a human rater was given, the less certain their interpretation became, signaling a hurdle for future automated emotion mining tools \cite{Murgia2014May}. Islam and Zibran conducted a large-scale quantitative study of over 490,000 commit comments from 50 open-source projects. Their findings showed emotional variations across different development activities and periods, adding to the understanding of emotions' roles and exposing opportunities for emotion-aware task assignment and collaboration \cite{Islam2016}. More recently, researchers have turned to social media to gauge emotional reactions to transformative technologies. Eshraghian et al. analyzed over 100,000 tweets about the AI programming assistant GitHub Copilot. They identified a spectrum of emotions, including challenge, loss, and skepticism, and observed a shift from negative to positive feelings as developers engaged with the tool and related its functionality to their professional identity \cite{Eshraghian2024Apr}.

Although Graziotin and Wrobel's work established a direct link between affect and productivity, the empirical quantification of this impact is underdeveloped. The complexity of establishing this connection is emphasized in experimental studies. In a controlled experiment, Anany attempted to use emotion and behavior monitoring to predict success in solving programming problems. However, the resulting classifier achieved only modest accuracy. This led the author to conclude that the work primarily ``opens a different track for experimentation,'' requiring more data and different methods~\cite{Anany2019}. Similarly, Girardi et al. conducted a field study that confirmed a positive correlation between emotional valence and perceived productivity. However, they found that their classifier, which was based on non-invasive biometric sensors, was not robust enough for practical use~\cite{Girardi2021Jun}. 

Broad surveys of the field also demonstrate these kinds of challenges. A review by Anany et al. concluded that the ``limited number of studies and the inconsistencies in the findings suggest the need for more effective ways to detect users' emotions and related productivity''~\cite{Anany2020Oct}. Likewise, another review highlighted that, although the beneficial effects of positive emotions on well-being are well-known, the specific mechanisms by which they improve developer performance require further investigation to enhance ergonomics and organizational health~\cite{Kurian2023Nov}. Overall, while a consensus is forming that emotions are critical, the literature indicates a clear and persistent need for more robust and validated methods to consistently link specific emotional states to measurable performance outcomes.

A significant methodological gap becomes noticeable when the existing literature is reviewed. While many studies draw on psychological theories or apply existing psychometric scales, there is a notable lack of research involving formal collaboration with psychology experts in designing and validating the measurement instruments. Emotions are complex, latent psychological constructs. Ensuring that survey questions accurately and comprehensively reflect the concept being measured requires in-depth knowledge of industrial-organizational psychology. Therefore, there is a clear need for research that bridges the interdisciplinary gap by grounding the measurement of developer emotions in a process validated by psychologists from the beginning. This approach is essential for building a reliable, valid foundation for understanding the relationship between emotions and developer productivity.

\section{Research framework and hypothesis formulation}
\label{sec:Research-framework-hypothesis-formulation}
The concepts of emotional state and perceived productivity are firmly established in social science research and practice. In fact, their status is rarely questioned. However, taking into account the context of our study, embedded in the domain of information systems, which relies heavily on the theory of human behavior, we need to explicitly define each factor by drawing on the state-of-the-art research. 

Since latent variables are theoretical concepts that cannot be directly observed, the first step is to explicitly and precisely define the concepts. Therefore, in the context of this study and prior theory, emotional state refers to a person's feelings and moods at a given moment in response to various stimuli \cite{iversen2000emotional, duque2013effects}. Perceived productivity, on the other hand, refers to an individual's sense of effectiveness in completing tasks or achieving goals within a specific period \cite{abdel1989software, aprilina2023role}. 

Emotions play a fundamental role in workplace productivity of a human by influencing collaboration \cite{huang2023social}, attention \cite{meyers2014emotional}, concentration \cite{sharma2018student}, and memory \cite{tyng2017influences}. In general, recent research shows employees' mental health and psychological state affect their productivity \cite{kadoya2020emotional}. In this vein, positive emotions stimulate self-motivation and increase the rate and capacity of proper information processing \cite{matuliauskaite2011analysis}.

The above applies to software developers who require creativity and problem-solving abilities in their day-to-day tasks. Key mental skills include adaptability \cite{liebenberg2014knowledge}, communication \cite{masood2022like}, critical thinking \cite{niva2023junior} and problem-solving \cite{booneka2011logical}. In practice, these skills are essential for addressing complex challenges, collaborating effectively, and implementing correct software solutions. While this type of work relies not only on technical expertise \cite{de2024aligning} and a range of cognitive skills \cite{mian2022comprehensive}, individual productivity is also affected by the emotions experienced \cite{Mendes2017}.

Since the link between emotions and productivity is well-recognized, the research background discussed above supports the following hypothesis:

\begin{itemize}
    \item[\textbf{H1.}] The emotional state of a software developer influences productivity.
\end{itemize}

\section{Methodology}
\label{sec:methodology}
In line with the objective of the undertaken research, we followed Standards and Guidelines for Validation discussed by Chan \cite{chan2014standards}. First, we transform the theoretical constructs into measurable indicators, which allows us to empirically investigate the relationship between them. Second, we submit the newly developed measurement instrument for expert review to assess whether its content is appropriate and valid. 

After validating the measurement instrument, we developed a questionnaire based on the guidelines provided by Patten \cite{patten2016questionnaire} and Jenn \cite{jenn2006designing} to collect data for hypothesis testing. To establish the reliability of the research, we present the respondent profile and justify both the chosen data analysis method and the software package used.

\subsection{Factors Operationalization}
By definition, an indicator is an observed variable that represents an unobserved (or latent) factor. The measurement items were developed based on relevant scholarly research by reputable authors. However, minor adjustments were made to align them with the specific research context. In this regard, the emotional state (ES) construct is based on the work of Aluoja et al. \cite{aluoja1999development}, Matthews et al.~\cite{matthews1999validation}, and Perkun et al.~\cite{pekrun2011measuring}.
Considering perceived productivity, our line of thinking was informed by the work of Baker et al.~\cite{baker2007satisfaction}, Gennara~\cite{gennara2023understanding}, Storey et al.~\cite{storey2019towards}, and Smite~\cite{smite2022changes}.

To sum up, in order to operationalize each factor, we developed an initial list of 40 indicators, including 20 items for emotional state (Table~\ref{tab:es-indicators-evaluation}) and 20 items for perceived productivity (Table~\ref{tab:pp-indicators-evaluation}). Since both minor and major adjustments were made to the original wording, it is necessary to validate the underlying logic and rationale behind all items.

\subsection{Validation}
The goal of validation is to confirm that the developed indicators accurately reflect the theoretical constructs of the two analyzed factors. In this regard, we designed and conducted a survey among experts in the field of research. To us, an expert is a person with specialized knowledge and experience in psychology. In this view, the requirements were twofold: first, educational level; and second, professional history.
Specifically, an expert must have at least a Master of Science degree in psychology or a related field and at least three years of professional experience.

We used a convenience sampling method to reach a group of experts \cite{stratton2021population}. This approach was chosen due to a lack of research funding and strict time constraints. We made initial contact via social media, telephone, or email. Given the specific nature of the research, we provided a brief description of the survey and its objectives. Participants were also assured anonymity and confidentiality. Additionally, no remuneration or incentives were offered. Between January and March of 2023, nine eligible respondents agreed to participate in the study.

The questionnaire contained 40 items written in both Polish and English. We see this as a fully justified action, considering the language of the future publication. Furthermore, respondents did not raise any issues with the terminology or the translation. The order of the items representing the two latent variables was mixed up. Each item was evaluated using a seven-point Likert scale, ranges from 1 denoting absolutely inappropriate, to 7, denoting absolutely appropriate. The other numbers on the scale were 2, inappropriate; 3, slightly inappropriate; 4, neutral; 5, slightly appropriate; and 6, appropriate. 

Table ~\ref{tab:es-indicators-evaluation} shows the results of the assessment of the emotional state variable carried out by a panel of experts. 

\begin{table}[H]
\centering
\small
\caption{A list of 20 indicators for the emotional state factor and the results of their evaluation by the experts ($n=9$).}
\label{tab:es-indicators-evaluation}
\begin{tabular}{|l|p{7.5cm}|l|l|l|l|}
\hline
Code  & Indicator                                                                     & Min & Max & AVG  & Score \\ \hline
ES1  & When I feel appreciated then I work more efficiently                          & 6   & 7   & 6.67 & 60    \\ \hline
ES2  & When I feel good about the work I am more productive                          & 5   & 7   & 6.44 & 58    \\ \hline
ES3  & Interest into the subject is what makes me to solve my task faster            & 5   & 7   & 6.33 & 57    \\ \hline
ES4  & Feeling enthusiastic allows me to   accomplish my tasks more efficiently      & 4   & 7   & 6.22 & 56    \\ \hline
ES5  & When I feel content my productivity is high                                   & 4   & 7   & 6.11 & 55    \\ \hline
ES6  & Feeling inspired is a driving force for me to be motivated to work            & 5   & 7   & 6.00 & 54    \\ \hline
ES7  & When I trust other team members, I am highly motivated to do a great job      & 4   & 7   & 5.89 & 53    \\ \hline
ES8  & Sense of direction/confidence drives me to continue my work                   & 4   & 7   & 5.89 & 53    \\ \hline
ES9  & When I am in love with what I am doing then I my work is most efficient       & 4   & 7   & 5.89 & 53    \\ \hline
ES10 & Enjoying my tasks makes me more efficient                                     & 3   & 7   & 5.78 & 52    \\ \hline
ES11 & While feeling delighted my productivity increases                             & 4   & 7   & 5.78 & 52    \\ \hline
ES12 & When I feel enthusiastic, I am more capable of solving problems               & 4   & 7   & 5.44 & 49    \\ \hline
ES13 & Being successful at work is what motivates me to do a good job                & 3   & 7   & 5.33 & 48    \\ \hline
ES14 & The more entertained I feel the faster I deal with my tasks                   & 3   & 6   & 5.33 & 48    \\ \hline
ES15 & Feeling happy makes me being motivated to work harder                         & 4   & 6   & 5.22 & 47    \\ \hline
ES16 & When I feel enjoyment it allows me to finish my work before the deadline      & 4   & 6   & 5.22 & 47    \\ \hline
ES17 & The more optimistic I feel the more efficient I am when solving the   problem & 3   & 6   & 5.11 & 46    \\ \hline
ES18 & Feeling secure makes me more efficient                                        & 3   & 7   & 5.11 & 46    \\ \hline
ES19 & When I feel excited then I am more productive                                 & 3   & 7   & 4.67 & 42    \\ \hline
ES20 & Feeling delightful makes me more capable of dealing with difficulties         & 4   & 6   & 4.56 & 41    \\ \hline
\end{tabular}
\end{table}

All of the developed items received scores above 4.50. The average score for the top six indicators was at least 6.00. In other words, on average, the panel of experts deemed these items to appropriately manifest the emotional state of the human subject. Note that a minimum of three observed variables (indicators) is generally recommended to represent a single latent variable. However, we chose those six indicators to operationalize the emotional state factor, which enables us to construct a more robust and testable model. 

Table~\ref{tab:pp-indicators-evaluation} presents the results of the assessment of the perceived productivity variable carried out by a panel of experts. 

\begin{table}[H]
\centering
\small
\caption{A list of 20 indicators for the perceived productivity factor and the results of their evaluation by the experts ($n=9$)}
\label{tab:pp-indicators-evaluation}
\begin{tabular}{|l|p{7.5cm}|l|l|l|l|}
\hline
Code  & Indicator  & Min & Max & AVG  & Score \\ \hline
PP1  & I intentionally organize my work to cultivate productivity so I can feel  content                                                                                           & 5   & 7   & 6.67 & 60    \\ \hline
PP2  & Accomplishing my professional tasks makes me feel pleased                                                                                                                    & 5   & 7   & 6.22 & 56    \\ \hline
PP3  & I solve more problems when I feel satisfied                                                                                                                                  & 5   & 7   & 6.22 & 56    \\ \hline
PP4  & I show initiative when I feel enthusiastic                                                                                                                                   & 5   & 7   & 6.22 & 56    \\ \hline
PP5  & My work is beneficial when I am pleased with what I am doing                                                                                                                 & 5   & 7   & 6.22 & 56    \\ \hline
PP6  & I create an order for the tasks that I complete and I can prioritize jobs   based on time and importance, which makes me optimistic that I will be able   to finish them all & 5   & 7   & 6.11 & 55    \\ \hline
PP7  & I am able to meet the tasks' objectives when I am pleased with what I am   doing                                                                                             & 3   & 7   & 5.78 & 52    \\ \hline
PP8  & I am more capable of doing my work when I feel appreciated                                                                                                                   & 4   & 7   & 5.78 & 52    \\ \hline
PP9  & I more eager to do my work while feeling passionate                                                                                                                          & 5   & 7   & 5.67 & 51    \\ \hline
PP10 & I close more tasks during the day when I feel pleased                                                                                                                        & 5   & 7   & 5.56 & 50    \\ \hline
PP11 & I work constructive as long as I am happy                                                                                                                                    & 3   & 7   & 5.56 & 50    \\ \hline
PP12 & I am more innovative while being optimistic                                                                                                                                  & 3   & 7   & 5.44 & 49    \\ \hline
PP13 & I start to work fruitful as soon as I feel delighted                                                                                                                         & 4   & 7   & 5.44 & 49    \\ \hline
PP14 & I become more prolific at my work when I am joyful                                                                                                                           & 4   & 7   & 5.33 & 48    \\ \hline
PP15 & I am more efficient when I feel content                                                                                                                                      & 4   & 6   & 5.11 & 46    \\ \hline
PP16 & My work is more profitable when I am feeling excited                                                                                                                         & 3   & 7   & 5.11 & 46    \\ \hline
PP17 & Being inventive goes along with me feeling enthusiastic                                                                                                                      & 3   & 7   & 5.00 & 45    \\ \hline
PP18 & I am more productive when I feel optimistic                                                                                                                                  & 3   & 6   & 4.89 & 44    \\ \hline
PP19 & A vision of problem-solving motivates me to further work on the subject                                                                                                      & 2   & 7   & 4.33 & 39    \\ \hline
PP20 & Being focused is guaranteed when I feel joy                                                                                                                                  & 3   & 7   & 4.11 & 37    \\ \hline
\end{tabular}
\end{table}

On average, all indicators received scores above 4.00. Similarly, the top first six indicators were selected to operationalize the perceived productivity factor. Furthermore, when compared with emotional state, it can be seen that experts evaluated these constructs more favorably.

In this regard, emotional state (ES) is the independent (exogenous) variable, whereas perceived productivity (PP) is the dependent (endogenous) variable. To verify the causal relationship between these variables, we employed a quantitative methodology and used a survey design, with a questionnaire as the instrument for data collection. 

\subsection{Data Collection}
By definition, a questionnaire is a survey research method involving a set of questions to gather results from interviewers. To reach a broad audience of respondents, the researchers conducted a mixed-mode survey that incorporated both online and traditional media. The traditional (paper-based) method allowed us to reach respondents directly, while the others were reached by mail, email, and telephone. Data were collected from April to May 2023. 

The developed questionnaire was written in Polish, and contained seventeen questions divided into two parts. The first part included five items regarding the following: (1) gender, (2) age, (3) professional experience in the IT industry, (4) level of education, and (5) current position. 
The second part included twelve statements which concerned independent variable (emotional state) and dependent variable (perceived productivity). In this regard, a respondent used a 7-point Likert scale to evaluate, ranging from 1 - Strongly disagree, 2 - Disagree, 3 - Somewhat disagree, 4 - Neither agree nor disagree, 5 - Somewhat agree, 6 - Agree, and 7 - Strongly agree. Note that the items were mixed in a random order.

\subsection{Sample profile}
Over the course of two months, a total of 156 responses were collected, with 88 respondents identifying as software developers. Only this subset will be considered for further analysis.

As shown in Table~\ref{tab:sample-profile}, 73 respondents (82.95\%) were men, 14 (15.91\%) were women, and one person (1.14\%) preferred not to disclose their gender. 
The respondents were divided into five age groups. The largest group was aged 18 to 25, comprising 35 individuals (39.77\%). Those aged 26 to 35 included 30 respondents (34.09\%), while the 36 to 45 age group, with 12 respondents (13.64\%), formed the third-largest group. The remaining 12.5\% were distributed between the 46 to 55 group (5.68\%) and those aged 56 or older (6.82\%). 
Two education levels dominated among the respondents: 46 (52.27\%) held a bachelor's degree and 31 (35.23\%) held a master's degree. The remaining 11 respondents (12.50\%) had completed only secondary school. 
Last but not least, in terms of professional experience, the largest group of respondents (35.23\%) had 3 to 5 years of experience. Next, 18.18\% of software developers reported having 10 to 20 years of experience, 14.77\% had 5 to 10 years, and 13.64\% had 1 to 2 years. The proportions of those with up to one year and those with 20 or more years of experience were equal, both at 9.09\%.

\begin{table}[H]
\small
\centering
\caption{Sample profile (\textit{n}=88).}
\label{tab:sample-profile}
\begin{tabular}{|l|l|r|r|}
\hline
Attribute                                & Value             & Frequency & Share (\%) \\ \hline
\multirow{3}{*}{Gender}                  & Male              & 73        & 82.95    \\ \cline{2-4} 
                                         & Female            & 14        & 15.91    \\ \cline{2-4} 
                                         & Prefer not to say & 1         & 1.14     \\ \hline
\multirow{5}{*}{Age}                     & 18--25             & 35        & 39.77    \\ \cline{2-4} 
                                         & 26--35             & 30        & 34.09    \\ \cline{2-4} 
                                         & 36--45             & 12        & 13.64    \\ \cline{2-4} 
                                         & 46--55             & 5         & 5.68     \\ \cline{2-4} 
                                         & 56 or more        & 6         & 6.82     \\ \hline
\multirow{3}{*}{Education}               & Bachelor          & 46        & 52.27    \\ \cline{2-4} 
                                         & Master of Science & 31        & 35.23    \\ \cline{2-4} 
                                         & Secondary         & 11        & 12.50    \\ \hline
\multirow{6}{*}{Professional Experience} & 0--1               & 8         & 9.09     \\ \cline{2-4} 
                                         & 1--2               & 12        & 13.64    \\ \cline{2-4} 
                                         & 3--5               & 31        & 35.23    \\ \cline{2-4} 
                                         & 5--10              & 13        & 14.77    \\ \cline{2-4} 
                                         & 10--20             & 16        & 18.18    \\ \cline{2-4} 
                                         & 20 or more        & 8         & 9.09     \\ \hline
\end{tabular}
\end{table}

\subsection{Data Analysis}
This study employs the structural equation modeling (SEM) approach, using Partial Least Squares (PLS) as the statistical method. PLS is a widely used technique for testing and validating cause-and-effect models \cite{adzgauskaite2025helps, kim2025online}. By design, it examines psychometric properties and provides evidence on the strength and direction of hypothesized relationships. To accomplish this, we conducted a two-stage data analysis following Anderson and Gerbing \cite{anderson1988structural} approach. The survey data were analyzed using SmartPLS software package (ver. 4.1.1.4) \cite{cheah2024reviewing}. 

\section{Results}
\label{sec:results}
This section is divided into two parts. In the first part, we evaluate the measurement model by assessing its reliability, convergent validity, and discriminant validity. In the second part, we evaluate the structural model. To this end, we test the hypothesized relationship between variables and evaluate the model's fit, focusing on its explanatory and predictive power.

\subsection{Measurement Model}
We begin evaluating the performance of the measurement model by analyzing its reliability (see Table~\ref{tab:reliability-convergent-validity}). We use Cronbach’s alpha ($\alpha$) to assess the internal consistency of the scales. A generally accepted rule of thumb is that the value of $\alpha$  for the construct should be equal to or greater than 0.70 \cite{almarzouqi2022prediction}. Both constructs included in the study's model demonstrate a high degree of internal consistency, with Cronbach's alpha values of 0.902 and 0.844 for emotional state (ES) and perceived productivity (PP), respectively. Composite reliability (CR) is another measure used to assess internal consistency. With values of 0.925 for ES and 0.885 for PP exceeding the commonly recommended threshold of 0.7 \cite{wang2006group}, the indicators reliably measure their intended constructs.

The second criterion evaluated is convergent validity, determined using the Average Variance Extracted (AVE) along with a detailed analysis of factor loadings \cite{almarzouqi2022prediction}. The AVE values of 0.675 for ES and 0.563 for PP both exceed the commonly recommended threshold of 0.5. Next, an evaluation of the factor loadings shows that two indicators, ES3 and PP6, have moderate loadings (greater than 0.5 but less than 0.7), while the remaining loadings exceed 0.7 and are considered high \cite{chang2011external}. Since all observed loadings are statistically significant, with p-values less than 0.001, this confirms their relevance to the underlying latent constructs. Together, these findings suggest strong convergent validity of the model, as shown in Table~\ref{tab:reliability-convergent-validity}.

\begin{table}[H]
\small
\centering
\caption{Results of reliability, convergent validity and statistical significance tests.}
\label{tab:reliability-convergent-validity}
\begin{tabular}{|l|l|c|c|c|c|c|}
\hline
Construct  & Item & Factor Loading & t-statistic & AVE   & CR    &  $\alpha$ \\ \hline
\multirow{6}{*}{Emotional State}        & ES1   & 0.828          & 14.525   & \multirow{6}{*}{0.675} & \multirow{6}{*}{0.925} & \multirow{6}{*}{0.902} \\ \cline{2-4}
                                        & ES2   & 0.887          & 27.101   &                        &                        &                        \\ \cline{2-4}
                                        & ES3   & 0.698          & 6.026   &                        &                        &                        \\ \cline{2-4}
                                        & ES4   & 0.849          & 21.424   &                        &                        &                        \\ \cline{2-4}
                                        & ES5   & 0.813          & 13.903   &                        &                        &                        \\ \cline{2-4}
                                        & ES6   & 0.841          & 19.643      &                        &                        &                        \\ \hline
\multirow{6}{*}{Perceived Productivity} & PP1   & 0.746          & 12.213   & \multirow{6}{*}{0.563} & \multirow{6}{*}{0.885} & \multirow{6}{*}{0.844} \\ \cline{2-4}
                                        & PP2   & 0.782          & 10.278   &                        &                        &                        \\ \cline{2-4}
                                        & PP3   & 0.794          & 14.148   &                        &                        &                        \\ \cline{2-4}
                                        & PP4   & 0.745          & 9.694   &                        &                        &                        \\ \cline{2-4}
                                        & PP5   & 0.773          & 10.141   &                        &                        &                        \\ \cline{2-4}
                                        & PP6   & 0.652          & 7.100   &                        &                        &                        \\ \hline
\end{tabular}
\end{table}

The third criterion is discriminant validity. Since establishing convergent validity is a prerequisite for discriminant validity, each indicator should load uniquely on only one construct. In practice, this means that an indicator's outer loading on its associated construct should be greater than any of its cross-loadings on other constructs. A careful reading of Table~\ref{tab:cross-loadings}, row by row, suggests that this assumption has been met for all indicators

\begin{table}[H]
\small
\centering
\caption{Discriminant validity - Cross loadings}
\label{tab:cross-loadings}
\begin{tabular}{|l|c|c|}
\hline
Indicator & Emotional State & Perceived Productivity \\ \hline
ES1       & \textbf{0.828}           & 0.720                  \\ \hline
ES2       & \textbf{0.887}           & 0.848                  \\ \hline
ES3       & \textbf{0.698}           & 0.600                  \\ \hline
ES4       & \textbf{0.849}           & 0.737                  \\ \hline
ES5       & \textbf{0.813}           & 0.755                  \\ \hline
ES6       & \textbf{0.841}           & 0.712                  \\ \hline
PP1       & 0.614           & \textbf{0.746}                  \\ \hline
PP2       & 0.695           & \textbf{0.782}                  \\ \hline
PP3       & 0.783           & \textbf{0.794}                  \\ \hline
PP4       & 0.670           & \textbf{0.745}                  \\ \hline
PP5       & 0.697           & \textbf{0.773}                  \\ \hline
PP6       & 0.520           & \textbf{0.652}                  \\ \hline
\end{tabular}
\end{table}

However, if one considers the Fornell–Larcker criterion and compares the square root of the AVE values (0.821 for ES and 0.750 for PP) with their inter-construct correlation of 0.893, it suggests that discriminant validity is not supported. However, some authors have recently criticized the Fornell–Larcker criterion, as it ignores the extent to which indicators load onto other constructs \cite{radomir2020discriminant}. 
Nevertheless, the Heterotrait–Monotrait ratio (HTMT) value of 1.008 also indicates poor discriminant validity, suggesting that the constructs may not be truly distinct. Note that, according to Ab Hamid et al. \cite{ab2017discriminant}, the HTMT criterion is considered a stringent measure.

\subsection{Structural Model}
The results of the path analysis show that, the emotional state of a software developer has a strong, positive, and significant impact ($\beta = 0.893$, $p < 0.001$) on perceived productivity, clearly confirming the formulated hypothesis (H1). 

The model fit can be evaluated by several measures. The central criterion for the structural model’s assessment is the coefficient of determination ($R^2$). The $R^2$ value of 0.797 of the variance in endogenous variable (PP) is explained by the exogenous variable (ES), meaning the model demonstrates a strong explanatory power. In addition, the effect size, measured by Cohen’s $f^2$, with a value of 3.924, indicates a very strong direct effect. 

Considering other measures, the SRMR (Standardized Root Mean Square Residual) value of 0.075 indicates a good fit ($< 0.08$), suggesting the model adequately captures the relationships between variables in the data \cite{mcneish2018thorny}.
While the Normed Fit Index (NFI) value of 0.827 ($\chi^2$ = 123.4) falls within the acceptable fit threshold (between 0.8 and 0.9) \cite{tasmin2008linking}, it indicates that the model's fit is close to optimal and reasonably consistent with the observed data. One can conclude that the model does not necessarily require further investigation or adjustments.

To assess the predictive power of the developed model, we use the Q-square ($Q^2$) statistic. The calculated $Q^2$ value of 0.796 indicates strong predictive relevance \cite{kadarsah2023role} and acceptable predictive accuracy, with an RMSE of 0.475. In other words, emotional state is a strong predictor of perceived productivity.

\newpage
\section{Discussion}
\label{sec:discussion}
Manifestly, the findings of this study are clear. In this study, we demonstrated that emotional state is a strong predictor of perceived productivity in the software development workplace. That said, it is necessary to further discuss the theoretical and practical implications of this conclusion, while also acknowledging its inherent limitations, stemming from the research design and settings.

\subsection{Theoretical implications}
The developed model determined the strength and direction of the relationship between emotional state and perceived productivity among software developers. In light of the obtained results, this study has demonstrated the strong and positive influence of emotions on perceived productivity, confirming the theoretically grounded connection between these two factors~\cite{Graziotin2014Mar}.
From a broader perspective, it supports previous research on psychology and organizational behavior by providing evidence that emotions significantly impact productivity in knowledge-intensive, cognitively demanding fields such as software engineering \cite{crawford2014influence}.

\subsection{Practical implications}
In our opinion, the results of this study demonstrate significant managerial implications for senior managers, team leaders, and individual software developers working on software products and services. 
Specifically, managers and team leaders should foster supportive and collaborative work environments to reduce emotional strain. Strategies such as flexible work schedules, balanced workloads, and regular check-ins can prevent burnout and help developers maintain emotional balance.

Second, at the project management level, we argue for the incorporation of emotion-aware practices. For example, monitoring emotional states during high-pressure phases and providing interventions such as mindfulness sessions, team-building exercises, or short breaks could enhance both individual well-being and productivity outcomes.

Third, at the organizational level, we recognize emotional well-being as a key driver of software developer productivity. Implementing programs focused on stress reduction, mental health support, and emotional resilience training can foster positive emotional states, ultimately leading to improved performance.

\subsection{Limitations}
While this research provides significant insights, it is important to acknowledge its limitations.
First, self-ratings of both variables, namely emotional state and perceived productivity, were collected at the same time, which could result in a consistency bias \cite{podsakoff1986self}. Second, self-ratings may be inflated due to the positive wording used in all measurement indicators, which could lead respondents to provide higher scores as a result of positive associations with the effective outcomes recalled \cite{kamoen2012positive}.

Third, one could argue that the sample size may significantly impact the accuracy of the results, preventing the findings from being generalized. However, considering the number of predictors and the recommended rule of five respondents per indicator \cite{moore2021data}, the sample size appears to satisfy both criteria \cite{goodhue2006pls}. On the other hand, as noted by Kahai and Cooper, one key advantage of PLS is its ability to produce reliable results with smaller sample sizes compared to other structural modeling approaches \cite{kahai2003exploring}.

\section{Conclusion}
\label{sec:conclusion}
Drawing on numerous theoretical arguments, we conclude that in all areas of human workplace activities, the role and effect of emotions is of utmost importance. In this line of thinking, our study found that a software developer’s emotional state is a strong and positive facilitator of perceived productivity.

Our research lays the groundwork for future emotion-centered studies in information system development, emphasizing the need for theoretical work that examines the mediating and moderating roles of emotions in relation to factors such as age, gender, grade level, job satisfaction, and team agility.

In this regard, future studies could employ longitudinal designs to capture changes in emotional state and productivity over time, providing deeper insights into causal relationships. Additionally, cross-cultural investigations may help to identify how cultural and organizational contexts influence the emotional dynamics of software developers. 

Ultimately, future research should examine how emotions are transferred at the team level, as well as the impact of collective emotional states on collaboration and decision-making both within and outside of teams.


\begin{thebibliography}{00}
\bibitem{keh1997classification}Keh, H. The classification of distribution channel output: a review. {\em The International Review Of Retail, Distribution And Consumer Research}. \textbf{7}, 145-156 (1997)

\bibitem{fulmer2014our}Fulmer, I. \& Ployhart, R. “Our Most Important Asset” a multidisciplinary/multilevel review of human capital valuation for research and practice. {\em Journal Of Management}. \textbf{40}, 161-192 (2014)

\bibitem{rajgopal2010mental}Rajgopal, T. Mental well-being at the workplace. {\em Indian Journal Of Occupational And Environmental Medicine}. \textbf{14} pp. 63-65 (2010)

\bibitem{dupuis1995examination}Dupuis, S. \& Smale, B. An examination of relationship between psychological well-being and depression and leisure activity participation among older adults. {\em Loisir Et Société/Society And Leisure}. \textbf{18}, 67-92 (1995)

\bibitem{bowling2012workload}Bowling, N. \& Kirkendall, C. Workload: A review of causes, consequences, and potential interventions. {\em Contemporary Occupational Health Psychology: Global Perspectives On Research And Practice, Volume 2}. \textbf{2} pp. 221-238 (2012)

\bibitem{zanabazar2022relationships}Zanabazar, A. \& Jigjiddorj, S. Relationships between mental workload, job burnout, and organizational commitment. {\em SHS Web Of Conferences}. \textbf{132} pp. 01003 (2022)

\bibitem{gaeta2020need}Gaeta, T. Need for a holistic approach to reducing burnout and promoting well-being. {\em Journal Of The American College Of Emergency Physicians Open}. \textbf{1}, 1050 (2020)

\bibitem{WHO2024}World Health Organization Mental health at work.  (2024), https://www.who.int/news-room/fact-sheets/detail/mental-health-at-work [Accessed on: 12.07.2025]

\bibitem{Taylor2020}Taylor, J. The Cost of Poor Workforce Mental Health.  (2020), https://sapienlabs.org/mentalog/the-cost-of-poor-mental-health-in-the-workplace/ [Accessed on: 12.07.2025]

\bibitem{WEF2025}World Economic Forum Mental Health.  (2025), https://intelligence.weforum.org/topics/a1Gb0000000pTDbEAM [Accessed on: 12.07.2025]

\bibitem{morris1997managing}Morris, J. \& Feldman, D. Managing emotions in the workplace. {\em Journal Of Managerial Issues}. pp. 257-274 (1997)

\bibitem{resendiz2021novel}Resendiz-Ochoa, E., Cruz-Albarran, I., Garduno-Ramon, M., Rodriguez-Medina, D., Osornio-Rios, R. \& Morales-Hernandez, L. Novel expert system to study human stress based on thermographic images. {\em Expert Systems With Applications}. \textbf{178} pp. 115024 (2021)

\bibitem{nawaz2024nexus}Nawaz, N., Gajenderan, V., Gopinath, U. \& Tharanya, V. Nexus between emotional intelligence and occupational stress: Role of workplace spirituality among teaching fraternity. {\em Asia Pacific Management Review}. \textbf{29}, 141-150 (2024)

\bibitem{yadav2025emotion}Yadav, G., Bokhari, M., Alzahrani, S., Alam, S. \& Shuaib, M. Emotion-aware ensemble learning (EAEL): revolutionizing Mental Health diagnosis of corporate professionals via Intelligent Integration of Multi-modal Data sources and ensemble techniques. {\em IEEE Access}. (2025)

\bibitem{barrett2006solving}Barrett, L. Solving the emotion paradox: Categorization and the experience of emotion. {\em Personality And Social Psychology Review}. \textbf{10}, 20-46 (2006)

\bibitem{kowal2025fire}Kowal, J. \& Winkler, E. Fire of Emotions: Challenges of Working Psychoanalytically in Extreme Times. {\em Jungian And Interdisciplinary Interfaces Between Emotions}. pp. 84-101 (2025)

\bibitem{kowal2024organizational}Kowal, J., Jasińska-Biliczak, A. \& Weichbroth, P. Organizational ethics and position relationship moderators among knowledge workers: a regional study of Poland. {\em Information Technology For Development}. pp. 1-23 (2024)

\bibitem{van2018under}Oorschot, K., Sengupta, K. \& Van Wassenhove, L. Under pressure: The effects of iteration lengths on agile software development performance. {\em Project Management Journal}. \textbf{49}, 78-102 (2018)

\bibitem{diamantopoulos2023semantically}Diamantopoulos, T., Nastos, D. \& Symeonidis, A. Semantically-enriched jira issue tracking data. {\em 2023 IEEE/ACM 20th International Conference On Mining Software Repositories (MSR)}. pp. 218-222 (2023)

\bibitem{kolimar2022delivery}Kolimar, O., Kusnirak, K., Kucera, E. \& Haffner, O. Delivery team management on small software development projects in practice. {\em 2022 Cybernetics \& Informatics (K\&I)}. pp. 1-6 (2022)

\bibitem{kaufmann2022does}Kaufmann, C. \& Kock, A. Does project management matter? The relationship between project management effort, complexity, and profitability. {\em International Journal Of Project Management}. \textbf{40}, 624-633 (2022)

\bibitem{bani2014collaborative}Bani-Salameh, H. \& Jeffery, C. Collaborative and social development environments: a literature review. {\em International Journal Of Computer Applications In Technology}. \textbf{49}, 89-103 (2014)

\bibitem{mishra2023structured}Mishra, A. \& Alzoubi, Y. Structured software development versus agile software development: a comparative analysis. {\em International Journal Of System Assurance Engineering And Management}. \textbf{14}, 1504-1522 (2023)

\bibitem{graziotin2017consequences}Graziotin, D., Fagerholm, F., Wang, X. \& Abrahamsson, P. Consequences of unhappiness while developing software. {\em 2017 IEEE/ACM 2nd International Workshop On Emotion Awareness In Software Engineering (SEmotion)}. pp. 42-47 (2017)

\bibitem{singh2012antecedents}Singh, P., Suar, D. \& Leiter, M. Antecedents, work-related consequences, and buffers of job burnout among Indian software developers. {\em Journal Of Leadership \& Organizational Studies}. \textbf{19}, 83-104 (2012)

\bibitem{palacio2022profession}Palacio, R., Cordova, G., Castro, L. \& Borrego, G. Profession-centric measures and indicators for occupational stress: An empirical study with novice software developers in Mexico. {\em IET Software}. \textbf{16}, 405-421 (2022)

\bibitem{meier2018stress}Meier, A., Kropp, M., Anslow, C. \& Biddle, R. Stress in agile software development: practices and outcomes. {\em International Conference On Agile Software Development}. pp. 259-266 (2018)

\bibitem{Gannon1979Dec}Gannon, J. Human Factors in Software Engineering. {\em Computer}. \textbf{12}, 6-7 (1979,12)

\bibitem{Boehm1988Oct}Boehm, B. \& Papaccio, P. Understanding and controlling software costs. {\em IEEE Transactions On Software Engineering}. \textbf{14}, 1462-1477 (1988,10)

\bibitem{weiss1996affective}Weiss, H. \& Cropanzano, R. Affective events theory. {\em Research In Organizational Behavior}. \textbf{18}, 1-74 (1996)

\bibitem{picard1997affective}Picard, R. Affective computing. {\em Guide Books}. (1997)

\bibitem{Shaw2004Apr}Shaw, T. The emotions of systems developers: an empirical study of affective events theory. {\em ACM Conferences}. pp. 124-126 (2004,4)

\bibitem{Wrobel2013}Wrobel, M. Emotions in the software development process. {\em 2013 6th International Conference On Human System Interactions (HSI)}. pp. 06-08 (2013)

\bibitem{Graziotin2013}Graziotin, D., Wang, X. \& Abrahamsson, P. Are Happy Developers More Productive?. {\em Product-Focused Software Process Improvement}. pp. 50-64 (2013)

\bibitem{Graziotin2014Mar}Graziotin, D., Wang, X. \& Abrahamsson, P. Happy software developers solve problems better: psychological measurements in empirical software engineering. {\em PeerJ}. \textbf{2} pp. e289 (2014,3)

\bibitem{Graziotin2015Jul}Graziotin, D., Wang, X. \& Abrahamsson, P. Do feelings matter? On the correlation of affects and the self-assessed productivity in software engineering. {\em Journal Of Software: Evolution And Process}. \textbf{27}, 467-487 (2015,7)

\bibitem{Graziotin2015Aug}Graziotin, D., Wang, X. \& Abrahamsson, P. How do you feel, developer? An explanatory theory of the impact of affects on programming performance. {\em PeerJ Computer Science}. \textbf{1} pp. e18 (2015,8)

\bibitem{Graziotin2015Sep}Graziotin, D., Wang, X. \& Abrahamsson, P. Understanding the affect of developers: theoretical background and guidelines for psychoempirical software engineering. {\em ACM Conferences}. pp. 25-32 (2015,9)

\bibitem{Graziotin2017Jun}Graziotin, D., Fagerholm, F., Wang, X. \& Abrahamsson, P. On the Unhappiness of Software Developers. {\em ACM Other Conferences}. pp. 324-333 (2017,6)

\bibitem{Graziotin2017May}Graziotin, D., Fagerholm, F., Wang, X. \& Abrahamsson, P. Unhappy developers: bad for themselves, bad for process, and bad for software product. {\em ACM Conferences}. pp. 362-364 (2017,5)

\bibitem{Graziotin2018Jun}Graziotin, D., Fagerholm, F., Wang, X. \& Abrahamsson, P. What happens when software developers are (un)happy. {\em Journal Of Systems And Software}. \textbf{140} pp. 32-47 (2018,6)

\bibitem{Araujo2025}Araufijo, A., Kalinowski, M., Paixao, M. \& Graziotin, D. Towards Emotionally Intelligent Software Engineers: Understanding Students' Self-Perceptions After a Cooperative Learning Experience. {\em 2025 IEEE/ACM 18th International Conference On Cooperative And Human Aspects Of Software Engineering (CHASE)}. pp. 27-28 (2025)

\bibitem{Murgia2014May}Murgia, A., Tourani, P., Adams, B. \& Ortu, M. Do developers feel emotions? an exploratory analysis of emotions in software artifacts. {\em ACM Conferences}. pp. 262-271 (2014,5)

\bibitem{Islam2016}Islam, M. \& Zibran, M. Towards understanding and exploiting developers' emotional variations in software engineering. {\em 2016 IEEE 14th International Conference On Software Engineering Research, Management And Applications (SERA)}. pp. 08-10 (2016)

\bibitem{Eshraghian2024Apr}Eshraghian, F., Hafezieh, N., Farivar, F. \& Cesare, S. AI in software programming: understanding emotional responses to GitHub Copilot. {\em Information Technology \& People}. \textbf{38}, 1659-1685 (2024,4)

\bibitem{Anany2019}Anany, M., Hussien, H., Aly, S. \& Sakr, N. Influence of Emotions on Software Developer Productivity. {\em Proceedings Of The 9th International Conference On Pervasive And Embedded Computing And Communication Systems - PECCS}. pp. 75-82 (2019)

\bibitem{Girardi2021Jun}Girardi, D., Lanubile, F., Novielli, N. \& Serebrenik, A. Emotions and Perceived Productivity of Software Developers at the Workplace. {\em IEEE Transactions On Software Engineering}. \textbf{48}, 3326-3341 (2021,6)

\bibitem{Anany2020Oct}Anany, M., Hussein, H. \& Aly, S. A survey on the influence of developer emotions on software coding productivity. {\em International Journal Of Social And Humanistic Computing}. (2020,10), https://www.inderscienceonline.com/doi/abs/10.1504/IJSHC.2020.111166

\bibitem{Kurian2023Nov}Kurian, R. \& Thomas, S. Importance of positive emotions in software developers' performance: a narrative review. {\em Theoretical Issues In Ergonomics Science}. (2023,11)

\bibitem{iversen2000emotional}Iversen, S., Kupfermann, I. \& Kandel, E. Emotional states and feelings. {\em Principles Of Neural Science}. \textbf{4} pp. 982-997 (2000)

\bibitem{duque2013effects}Duque, M., Turla, C. \& Evangelista, L. Effects of emotional state on decision making time. {\em Procedia-Social And Behavioral Sciences}. \textbf{97} pp. 137-146 (2013)

\bibitem{abdel1989software}Abdel-Hamid, T. \& Madnick, S. Software productivity: potential, actual, and perceived. {\em System Dynamics Review}. \textbf{5}, 93-113 (1989)

\bibitem{aprilina2023role}Aprilina, R. \& Martdianty, F. The Role of Hybrid-Working in Improving Employees' Satisfaction, Perceived Productivity, and Organizations' Capabilities.. {\em Jurnal Manajemen Teori Dan Terapan}. \textbf{16} (2023)

\bibitem{huang2023social}Huang, X. \& Lajoie, S. Social emotional interaction in collaborative learning: Why it matters and how can we measure it?. {\em Social Sciences \& Humanities Open}. \textbf{7}, 100447 (2023)

\bibitem{meyers2014emotional}Meyers, J., Grills, C., Zellinger, M. \& Miller, R. Emotional distress affects attention and concentration: the difference between mountains and valleys. {\em Applied Neuropsychology: Adult}. \textbf{21}, 28-35 (2014)

\bibitem{sharma2018student}Sharma, P., Esengönül, M., Khanal, S., Khanal, T., Filipe, V. \& Reis, M. Student concentration evaluation index in an e-learning context using facial emotion analysis. {\em International Conference On Technology And Innovation In Learning, Teaching And Education}. pp. 529-538 (2018)

\bibitem{tyng2017influences}Tyng, C., Amin, H., Saad, M. \& Malik, A. The influences of emotion on learning and memory. {\em Frontiers In Psychology}. \textbf{8} pp. 235933 (2017)

\bibitem{kadoya2020emotional}Kadoya, Y., Khan, M., Watanapongvanich, S. \& Binnagan, P. Emotional status and productivity: Evidence from the special economic zone in Laos. {\em Sustainability}. \textbf{12}, 1544 (2020)

\bibitem{matuliauskaite2011analysis}Matuliauskaitė, A. \& Žemeckytė, L. Analysis of interdependencies between students’ emotions, learning productivity, academic achievements and physiological parameters. {\em Mokslas–Lietuvos Ateitis/Science–Future Of Lithuania}. \textbf{3}, 51-56 (2011)

\bibitem{liebenberg2014knowledge}Liebenberg, J., Huisman, M. \& Mentz, E. Knowledge and skills requirements for software developer students. {\em International Journal Of Social, Behavioral, Educational, Economic, Business And Industrial Engineering}. \textbf{8}, 2604-2609 (2014)

\bibitem{masood2022like}Masood, Z., Hoda, R., Blincoe, K. \& Damian, D. Like, dislike, or just do it? How developers approach software development tasks. {\em Information And Software Technology}. \textbf{150} pp. 106963 (2022)

\bibitem{niva2023junior}Niva, A., Markkula, J. \& Annanperä, E. Junior software engineers’ international communication and collaboration competences. {\em IEEE Access}. \textbf{11} pp. 139039-139068 (2023)

\bibitem{booneka2011logical}Booneka, N. \& Kiattikomol, P. Logical Thinking Skills of Software Developers. {\em The Social Sciences}. \textbf{6}, 495-501 (2011)

\bibitem{de2024aligning}Campos, V., David, J., Ströele, V. \& Braga, R. Aligning technical knowledge to an industry domain in global software development: A systematic mapping. {\em Journal Of Software: Evolution And Process}. \textbf{36}, e2713 (2024)

\bibitem{mian2022comprehensive}Mian, I., Anwar, A., Alroobaea, R., Ullah, S., Almansour, F. \& Umar, F. A comprehensive skills analysis of novice software developers working in the professional software development industry. {\em Complexity}. \textbf{2022}, 2631727 (2022)

\bibitem{Mendes2017}Mendes, N. How do emotions affect productivity? [New research].  (2017), https://www.atlassian.com/blog/software-teams/new-research-emotional-intelligence-in-the-workplace [Accessed on: 03.08.2025]

\bibitem{chan2014standards}Chan, E. Standards and guidelines for validation practices: Development and evaluation of measurement instruments. {\em Validity And Validation In Social, Behavioral, And Health Sciences}. pp. 9-24 (2014)

\bibitem{patten2016questionnaire}Patten, M. Questionnaire research: A practical guide. (routledge,2016)

\bibitem{jenn2006designing}Jenn, N. Designing a questionnaire. {\em Malaysian Family Physician: The Official Journal Of The Academy Of Family Physicians Of Malaysia}. \textbf{1}, 32 (2006)

\bibitem{aluoja1999development}Aluoja, A., Shlik, J., Vasar, V., Luuk, K. \& Leinsalu, M. Development and psychometric properties of the Emotional State Questionnaire, a self-report questionnaire for depression and anxiety. {\em Nordic Journal Of Psychiatry}. \textbf{53}, 443-449 (1999)

\bibitem{matthews1999validation}Matthews, G., Joyner, L., Gilliland, K., Campbell, S., Falconer, S., Huggins, J. \& Others Validation of a comprehensive stress state questionnaire: Towards a state big three. {\em Personality Psychology In Europe}. \textbf{7} pp. 335-350 (1999)

\bibitem{pekrun2011measuring}Pekrun, R., Goetz, T., Frenzel, A., Barchfeld, P. \& Perry, R. Measuring emotions in students’ learning and performance: The Achievement Emotions Questionnaire (AEQ). {\em Contemporary Educational Psychology}. \textbf{36}, 36-48 (2011)

\bibitem{baker2007satisfaction}Baker, E., Avery, G. \& Crawford, J. Satisfaction and perceived productivity when professionals work from home. {\em Research \& Practice In Human Resource Management}. (2007)

\bibitem{gennara2023understanding}Gennara, A. Understanding and Measuring Perceived Productivity. (Carleton University,2023)

\bibitem{storey2019towards}Storey, M., Zimmermann, T., Bird, C., Czerwonka, J., Murphy, B. \& Kalliamvakou, E. Towards a theory of software developer job satisfaction and perceived productivity. {\em IEEE Transactions On Software Engineering}. \textbf{47}, 2125-2142 (2019)

\bibitem{smite2022changes}Smite, D., Tkalich, A., Moe, N., Papatheocharous, E., Klotins, E. \& Buvik, M. Changes in perceived productivity of software engineers during COVID-19 pandemic: The voice of evidence. {\em Journal Of Systems And Software}. \textbf{186} pp. 111197 (2022)

\bibitem{stratton2021population}Stratton, S. Population research: convenience sampling strategies. {\em Prehospital And Disaster Medicine}. \textbf{36}, 373-374 (2021)

\bibitem{adzgauskaite2025helps}Adzgauskaite, M., Tam, C. \& Martins, R. What helps Agile remote teams to be successful in developing software? Empirical evidence. {\em Information And Software Technology}. \textbf{177} pp. 107593 (2025)

\bibitem{kim2025online}Kim, C., Lee, J. \& Lee, K. Online review data analytics to explore factors affecting consumers’ airport recommendations. {\em Information Technology \& People}. \textbf{38}, 1890-1924 (2025)

\bibitem{anderson1988structural}Anderson, J. \& Gerbing, D. Structural equation modeling in practice: A review and recommended two-step approach.. {\em Psychological Bulletin}. \textbf{103}, 411 (1988)

\bibitem{cheah2024reviewing}Cheah, J., Magno, F. \& Cassia, F. Reviewing the SmartPLS 4 software: the latest features and enhancements. (Springer,2024)

\bibitem{almarzouqi2022prediction}Almarzouqi, A., Aburayya, A. \& Salloum, S. Prediction of user’s intention to use metaverse system in medical education: A hybrid SEM-ML learning approach. {\em IEEE Access}. \textbf{10} pp. 43421-43434 (2022)

\bibitem{wang2006group}Wang, E., Ying, T., Jiang, J. \& Klein, G. Group cohesion in organizational innovation: An empirical examination of ERP implementation. {\em Information And Software Technology}. \textbf{48}, 235-244 (2006)

\bibitem{chang2011external}Chang, K., Wong, J., Li, Y., Lin, Y. \& Chen, H. External social capital and information systems development team flexibility. {\em Information And Software Technology}. \textbf{53}, 592-600 (2011)

\bibitem{radomir2020discriminant}Radomir, L. \& Moisescu, O. Discriminant validity of the customer-based corporate reputation scale: Some causes for concern. {\em Journal Of Product \& Brand Management}. \textbf{29}, 457-469 (2020)

\bibitem{ab2017discriminant}Ab Hamid, M., Sami, W. \& Sidek, M. Discriminant validity assessment: Use of Fornell \& Larcker criterion versus HTMT criterion. {\em Journal Of Physics: Conference Series}. \textbf{890}, 012163 (2017)

\bibitem{mcneish2018thorny}McNeish, D., An, J. \& Hancock, G. The thorny relation between measurement quality and fit index cutoffs in latent variable models. {\em Journal Of Personality Assessment}. \textbf{100}, 43-52 (2018)

\bibitem{tasmin2008linking}Tasmin, R. \& Woods, P. Linking knowledge management and innovation: A structural equation modeling approach. {\em Innovation And Knowledge Management In Business Globalization: Theory \& Practice}. \textbf{1}, 558-65 (2008)

\bibitem{kadarsah2023role}Kadarsah, D., Govindaraju, R. \& Prihartono, B. The role of knowledge-oriented leadership in fostering innovation capabilities: the mediating role of data analytics maturity. {\em IEEE Access}. \textbf{11} pp. 129683-129702 (2023)

\bibitem{graziotin2014happy}Graziotin, D., Wang, X. \& Abrahamsson, P. Happy software developers solve problems better: psychological measurements in empirical software engineering. {\em PeerJ}. \textbf{2} pp. e289 (2014)

\bibitem{crawford2014influence}Crawford, B., Soto, R., Barra, C., Crawford, K. \& Olguin, E. The influence of emotions on productivity in software engineering. {\em International Conference On Human-Computer Interaction}. pp. 307-310 (2014)

\bibitem{podsakoff1986self}Podsakoff, P. \& Organ, D. Self-reports in organizational research: Problems and prospects. {\em Journal Of Management}. \textbf{12}, 531-544 (1986)

\bibitem{kamoen2012positive}Kamoen, N. Positive versus negative: A cognitive perspective on wording effects for contrastive questions in attitude surveys.  (2012)

\bibitem{moore2021data}Moore, Z., Harrison, D. \& Hair, J. Data quality assurance begins before data collection and never ends: What marketing researchers absolutely need to remember. {\em International Journal Of Market Research}. \textbf{63}, 693-714 (2021)

\bibitem{goodhue2006pls}Goodhue, D., Lewis, W. \& Thompson, R. PLS, small sample size, and statistical power in MIS research. {\em Proceedings Of The 39th Annual Hawaii International Conference On System Sciences (HICSS'06)}. \textbf{8} pp. 202b-202b (2006)

\bibitem{kahai2003exploring}Kahai, S. \& Cooper, R. Exploring the core concepts of media richness theory: The impact of cue multiplicity and feedback immediacy on decision quality. {\em Journal Of Management Information Systems}. \textbf{20}, 263-299 (2003)

\end{thebibliography}

\end{document}